\documentclass[twocolumn,prl,aps,showpacs,superscriptaddress]{revtex4}

\usepackage[dvips]{color}
\usepackage[english]{babel}
\usepackage{latexsym}

\usepackage{graphicx} 
\usepackage{epsf} 
\usepackage{subfigure} 
\usepackage{amsmath} 
\usepackage{amssymb} 
\usepackage{amsfonts}
\usepackage{bm}
\usepackage{natbib}
\usepackage{epstopdf}
\usepackage{appendix}

\definecolor{mygrey}{gray}{0.35}
\definecolor{mygreen}{rgb}{0.85,1,0.9}
\definecolor{myzard}{cmyk}{0,0,0.05,0}
\definecolor{mywhite}{rgb}{1,1,1}
\definecolor{myred}{rgb}{1,0,0}

 \def\ee{\mathord{\rm e}}
 
 \def\ii{\mathord{\rm i}}

\def\half{\textstyle\frac{1}{2}}

\def\bs#1{\boldsymbol{#1}}
\def\txt#1{\textrm{#1}}
\def\be{\begin{equation}}
\def\ee{\end{equation}}

\renewcommand{\ii}{{\rm i}}
\renewcommand{\ee}{{\rm e}}
\newcommand{\alf}{\Phi_\alpha}
\newcommand{\bet}{\Phi_\beta}
\newcommand{\ket}[1]{|#1\rangle}
\newcommand{\bra}[1]{\langle #1|}

\begin{document}

\title[Short Title]{Non-Abelian optical lattices: Anomalous quantum Hall effect and Dirac fermions}

\author{
N. Goldman
}
\affiliation{Center for Nonlinear Phenomena and Complex Systems - Universit$\acute{e}$ Libre de Bruxelles (U.L.B.), Code Postal 231, Campus Plaine, B-1050 Brussels, Belgium}

\author{A. Kubasiak 
}
\affiliation{
ICFO-Institut de Ci\`encies Fot\`oniques,
Parc Mediterrani de la Tecnologia,
E-08860 Castelldefels (Barcelona), Spain}

\affiliation{Marian Smoluchowski Institute of Physics Jagiellonian University, Reymonta 4, 30059 Krak\'ow, Polska
}

\author{A. Bermudez 
}
\affiliation{
Departamento de F\'isica Te\'orica I,
Universidad Complutense, 
28040 Madrid, 
Spain
}

\author{
P. Gaspard
}
\affiliation{Center for Nonlinear Phenomena and Complex Systems - Universit$\acute{e}$ Libre de Bruxelles (U.L.B.), Code Postal 231, Campus Plaine, B-1050 Brussels, Belgium}

\author{ 
M. Lewenstein}

\affiliation{
ICFO-Institut de Ci\`encies Fot\`oniques,
Parc Mediterrani de la Tecnologia,
E-08860 Castelldefels (Barcelona), Spain}
\affiliation{
ICREA - Instituci\`o Catalana de Ricerca i Estudis Avan{\c c}ats, 08010 
Barcelona, Spain}
\author{M.A. Martin-Delgado}

\affiliation{
Departamento de F\'isica Te\'orica I,
Universidad Complutense, 
28040 Madrid, 
Spain
}


\begin{abstract}
We study the properties of an ultracold Fermi gas loaded in an  optical square lattice and subjected to an external and classical non-Abelian gauge field. We show that this system can be exploited as an optical analogue of relativistic quantum electrodynamics, offering a remarkable route to access the exotic properties of massless Dirac fermions with cold atoms experiments. In particular we show that  the underlying  Minkowski space-time can also be modified, reaching anisotropic regimes where a remarkable anomalous quantum Hall effect and a squeezed Landau vacuum could be observed.
\end{abstract}

\maketitle

Low energy excitations of fermionic lattice systems are usually governed by the non-relativistic Schr\"odinger equation. However, this description must be profoundly altered in the vicinity of 
Dirac points, where the energy bands display conical singularities  and quasiparticles become massless relativistic fermions. Such a remarkable behavior can be induced by a honeycomb geometry \cite{semenoff,graphene_review,nature,nature2,duan}, or by additional uniform~\cite{wen} or staggered~\cite{emmerich,recent} magnetic fields. Here we show that the natural playground for emerging Dirac fermions is provided by multi-component fermionic atoms subjected to artificial 
non-Abelian gauge fields. We emphasize that these external fields can be produced by generalizing  the recent experiment~\cite{spielman}, as proposed in~\cite{osterloh,ohberg}. Such gauge fields  give rise to intriguing phenomena such as the non-Abelian Aharonov-Bohm effect~\cite{osterloh},  generation of magnetic monopoles~\cite{monopole}, non-Abelian atom optics~\cite{neg_rafraction}, quasi-relativistic effects~\cite{ohberg2}, or even the modification of the metal-insulator transition~\cite{metal_insulator}. In this Letter, we show that  the physical properties of  massless relativistic fermions are completely characterized by the non-Abelian features of the external gauge fields. Furthermore the anisotropy  of the underlying Minkowski space-time can be controlled externally,  producing an anomalous quantum  Hall effect characterized by a squeezed Landau vacuum. 

\begin{center} 
\begin{figure}
\begin{center}
\hspace{-0.1cm}{\scalebox{0.13}{\includegraphics{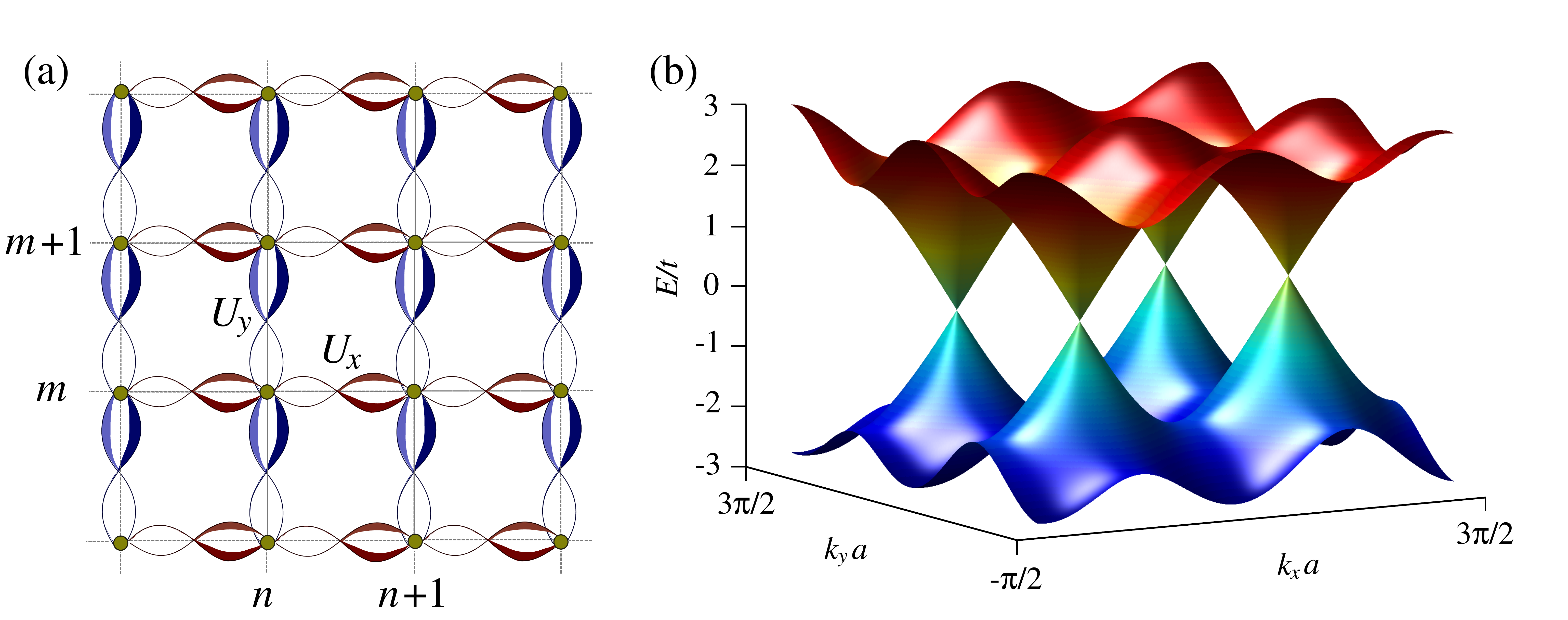}}}
\caption{\label{square_lattice} (a) Square lattice subjected to a non-Abelian gauge potential. This external field  induces state-dependent hoppings described by the U(2) operators $U_{x}$ and $U_y$. (b) Energy bands  close to the $\pi$-flux regime ($\Phi_{\alpha}=\pi/2+0.1$, $\Phi_{\beta}=\pi/2-0.1$), with vanishing Abelian flux $\Phi=0$. The bands touch at four Dirac points inside the first Brillouin zone (BZ), where the energy scales linearly with momenta $E\sim k$.} 
\end{center}
\end{figure} 
\end{center}

We consider a system of  two-component (two-color) fermionic atoms trapped in an  optical square lattice with sites at $\textbf{r}=(n,m)a$, where $a$ is the lattice spacing and $n,m\in\mathbb{Z}$. In the non-interacting limit, which can be obtained by means of  Feshbach resonances \cite{feshbach}, fermions freely hop between neighboring   sites. The addition of an external gauge potential $\textbf{A}$  modifies the hopping Hamiltonian according to the Peierls substitution
\begin{equation}
H=-t\sum_{\langle \textbf{r},\textbf{r}'\rangle}\sum_{\tau\tau'}c_{\tau'}^{\dagger}(\textbf{r}')  \, \ee^{- \ii \int_{\textbf{r}}^{\textbf{r}'} \textbf{A} \cdot \textbf{dl}}  \, c_{\tau}(\textbf{r}) +\text{h.c.},
\end{equation}
where $t$ is the hopping amplitude, $ c_{\tau}(\textbf{r})$ is the fermionic field operator in color component $\tau=1,2$, and we set $\hbar=e=1$. Our setup features an  external gauge potential  with both commutative and non-commutative  components $\textbf{A}=\textstyle{\frac{B_0}{2}}(-y,x)+a(B_{\alpha} \sigma_y,B_{\beta} \sigma_x)$, where $B_0,B_{\alpha},B_{\beta}$ are controllable parameters and $\sigma_{x,y}$ are Pauli matrices. Accordingly the hoppings are accompanied by non-trivial unitary operators, $U_{x}(m)=\ee^{-\ii\tiny{\pi \Phi m}}\ee^{\ii \Phi_\alpha \sigma_y}$ and $U_y(n)=\ee^{\ii\tiny{\pi \Phi n}}\ee^{\ii \Phi_\beta \sigma_x}$, where $\Phi=\textstyle{B_0 a^2}$ is  the Abelian magnetic flux  and $\Phi_{\alpha,\beta}=\textstyle{B_{\alpha,\beta}a^2}$ are  the non-Abelian fluxes  (see Fig.~\ref{square_lattice}.a). 

Let us point out that the gauge fields considered in this work can be realized following the proposals~ \cite{osterloh,ohberg,goldman}, along the lines of the recent experiment~ \cite{spielman}, and provide non-Abelian analogues of 
homogeneous magnetic fields  since they are characterized by constant Wilson loops. Indeed, atoms hopping around an elementary plaquette undergo a unitary transformation $U=U_x (m) U_y(n + 1)U_x^{\dagger} (m+1) U^{\dagger}_y(n)$, explicitly given by
\begin{align}
&U=\ee^{ \ii 2\pi\Phi}\left(c_1 \mathbb{I} +c_2\sigma_z 
+c_3\sigma_y+c_4\sigma_x\right), 
\label{wilson}
\end{align}
where the constants $\{c_j\}$ are listed in \cite{constants}. For specific values of $\Phi_{\alpha,\beta}$ the loop matrix reduces to a phase factor and reproduces the Abelian $\pi$-flux ($\Phi_{\alpha}=\Phi_{\beta}=\frac{\pi}{2}$) or  Hofstadter ($\Phi_{\alpha}=\Phi_{\beta}=0$) models \cite{hofstadter,wen}. However in general cases, it is a non-trivial U(2) operator exhibiting non-Abelian properties such as the non-Abelian Aharonov-Bohm effect. The gauge-invariant Wilson loop $W=\text{tr}\, U$ provides a clear distinction between the Abelian ($\vert W \vert=2$) and non-Abelian ($\vert W \vert<2$) regimes. We stress that the Wilson loop is homogeneous and that the corresponding spectrum exhibits well developed gaps \cite{goldman}.

\begin{center} 
\begin{figure}
\begin{center}
\hspace{-0.cm}{\scalebox{0.13}{\includegraphics{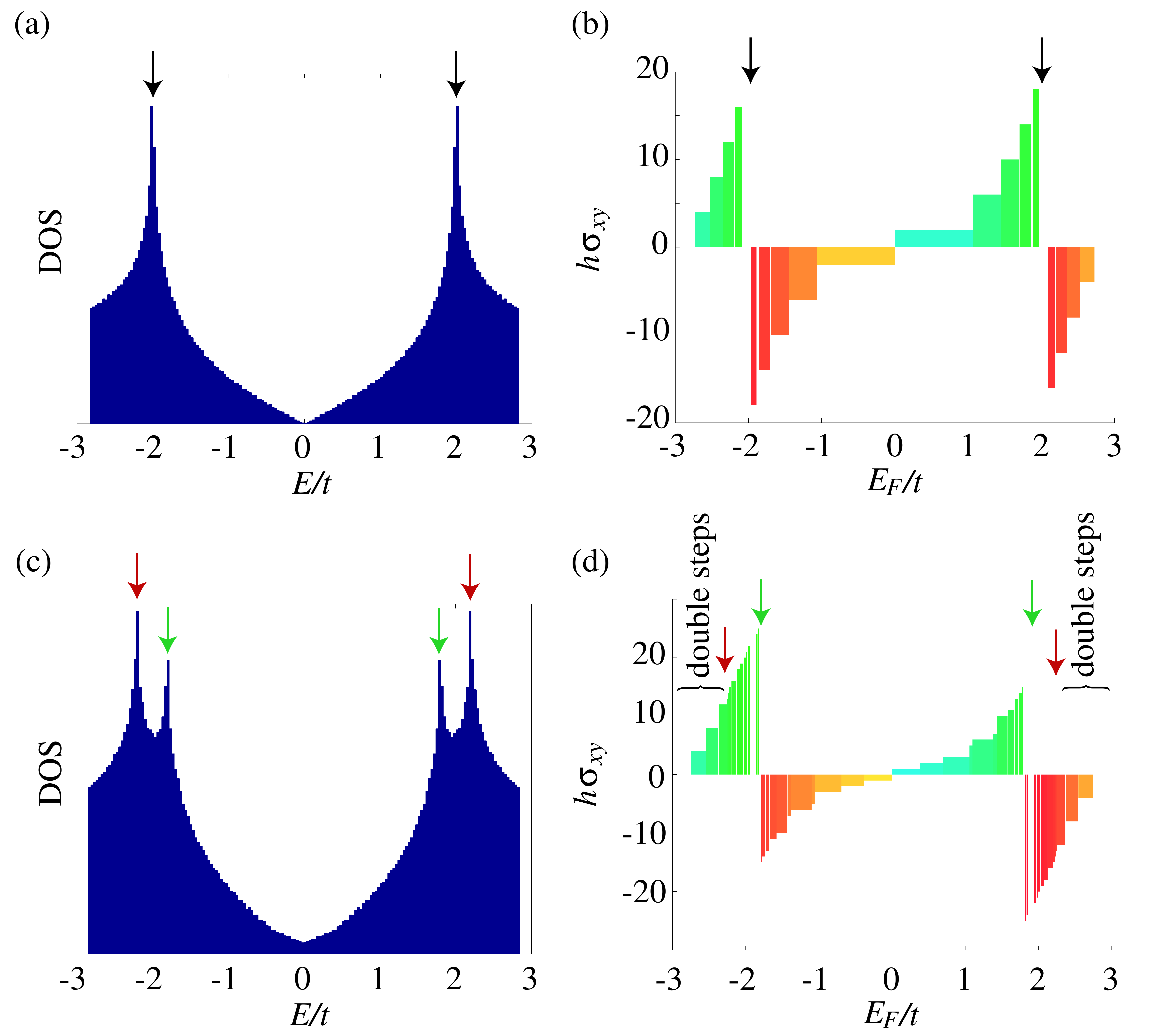}}} 
\caption{\label{thefig2} (a) Density of states (DOS) in the $\pi$-flux regime $\Phi_\alpha=\Phi_\beta=\pi/2$ when $\Phi=0$. (b) Hall conductivity in units of $h^{-1}$ as a function of the Fermi energy in the same regime  for $\Phi=1/41$. Black arrows designate the VHS. (c) DOS close to the $\pi$-flux regime $\Phi_\alpha=\pi/2+0.1$ and $\Phi_\beta=\pi/2-0.1$ when $\Phi=0$. (d) Hall conductivity $h \sigma_{xy}= h \sigma_{xy} (E_F)$ in the same regime as (c) for $\Phi=1/41$. Dark red and light green arrows respectively designate the VHS $E^{\txt{VHS}}_{\txt{red}}$ and $E^{\txt{VHS}}_{\txt{green}}$ (cf. Eq. \eqref{vhs}).} 
\end{center}
\end{figure} 
\end{center}

In order to isolate non-Abelian effects, we first study the regime of vanishing Abelian flux $\Phi=0$. The Hamiltonian is diagonalized  in momentum space and the fermion gas becomes a collection of non-interacting quasi-particles with  energies shown in  Fig.\ref{square_lattice}.b. Close to the marginally Abelian regime ($\Phi_{\alpha},\Phi_{\beta} \approx \pi/2$), the spectrum develops four independent conical singularities $\textbf{k}_{\text{D}}\in\{(0,0),(\frac{\pi}{a},0),(0,\frac{\pi}{a}),(\frac{\pi}{a},\frac{\pi}{a})\}\in {\text{BZ}}$, which correspond to massless relativistic excitations at half filling. Around these points $\textbf{p}=\textbf{k}-\textbf{k}_{\text{D}}$, the low-energy properties are accurately described by a  Dirac Hamiltonian
\begin{equation}
\label{dirac_equation}
H_{\text{eff}}=\sum_{\textbf{p}}{\Psi}_{\textbf{p}}^{\dagger}
H_{\text{D}} {\Psi}_{\textbf{p}},\hspace{3ex} H_{\text{D}} =c_{x}\alpha_xp_x+c_{y}\alpha_yp_y,
\end{equation}
 where  ${\Psi}_{\textbf{p}}=(c_{1\textbf{p}}, c_{2\textbf{p}})^{t}$ is the relativistic spinor, the Dirac matrices $\alpha_x,\alpha_y$ fulfill  
 $\{\alpha_j,\alpha_k\}=2\delta_{jk}$ (e.g. around $\textbf{k}_{\text{D}}=(0,\pi /a)$, $\alpha_x=\sigma_y$ and $\alpha_y=\sigma_x$), and  $c_{x}= 2 a t \sin\Phi_{\alpha}$, $c_y= 2 a t \sin{\Phi_{\beta}}$ represent the effective speed of light. We stress here that the control over the non-Abelian fluxes $\Phi_{\alpha,\beta}$ offers the exotic opportunity to modify the structure of the underlying Minkowski space-time, reaching anisotropic situations where $c_x\neq c_y$. Hence, non-Abelian optical lattices provide a quantum optical  analogue of relativistic QED,  where  the emerging  fermions and the properties of the corresponding space-time rely on the non-Abelian features of the external fields. Furthermore, it is also possible to observe a transition between relativistic and non-relativistic dispersion relations as the energy is increased. This abrupt change of the quasi-particle nature is revealed by Van Hove singularities (VHS) in the density of states, as displayed in  Figs.\ref{thefig2}.a-c. 

%Transport properties of two-dimensional Fermi gases subjected to orthogonal electric and magnetic fields  provide a paradigmatic scenario where to test the departure between quantum and classical predictions. Defying classical laws, the system's transverse conductivity takes on precise quantized values  $\sigma_{xy}=\frac{\nu}{h}$ with $\nu\in\mathbb{Z}$ , which  can  only be reconciled with a quantum theory of the Hall effect (QHE) and serve as a precision measurement of fundamental constants~\cite{hall_book}. As shown below,
%non-Abelian effects have dramatic consequences on the optical lattice analogue of the QHE, which occurs in presence of an additional Abelian magnetic field $B_0$ and an electric field, the latter being simulated by accelerating the  lattice. 

The transport properties of 2D Fermi gases subjected to external gauge fields are characterized by the optical-lattice analogue of the well-known quantum Hall effect (QHE) \cite{hall_book}. In this context, the transverse Hall conductivity measures the response of the system to a static force, e.g. a lattice acceleration, and takes on quantized values $\sigma_{xy}=\frac{\nu}{h}$ with $\nu\in\mathbb{Z}$, when the Fermi energy $E_F$ lies in a gap \cite{goldman}. Surprinsingly, the quantized conductivity of cold gases can be directly observed through density measurements thanks to the Streda formula \cite{oktel}. Here we show that non-Abelian effects have dramatic consequences on the QHE which occurs when an additional Abelian flux $\Phi$ is applied to our system.
The quantized values of the transverse conductivity are calculated as the sum of topological invariants associated to each energy band, the so-called Chern numbers~\cite{Kohmoto1985},
\begin{align}
&\sigma_{xy}= -  \sum_{E_n<E_F} \frac{i}{2 \pi h}\int_{BZ} \txt{tr} \, \mathcal{F}(\psi_n) \, \bs{\txt{\bf{d}} k}, 
\label{chern}
\end{align}
where $\mathcal{F}(\psi_n)=\langle \partial_{k_{x}} \psi_n \vert \partial_{k_{y}} \psi_n \rangle-\langle \partial_{k_{y}} \psi_n \vert \partial_{k_{x}} \psi_n \rangle $ is the Berry's curvature of the band $E_n$. Here the Chern numbers are computed numerically by discretizing the Brillouin zone \cite{Fukui2005}. A lattice gauge theory method allows to determine the Berry's curvature
 \begin{align}
&\mathcal{F}_{xy} (\boldsymbol{k}_l)= \textrm{ln} \, T_x (\boldsymbol{k}_l) T_y (\boldsymbol{k}_l +\hat{ \boldsymbol{x}}) T_x (\boldsymbol{k}_l +\hat{ \boldsymbol{y}})^{-1} T_y (\boldsymbol{k}_l)^{-1} , \notag \\
&T_{\mu} (\boldsymbol{k}_l)= \langle \psi_n (\boldsymbol{k}_l) \vert \psi_n (\boldsymbol{k}_l+\hat{ \boldsymbol{\mu}}) \rangle ,
\end{align}
and  subsequently the Chern number $C=\frac{i}{2 \pi } \sum_l \mathcal{F}_{xy} (\boldsymbol{k}_l)$. Remarkably, the sequence of Hall plateaus is extremely sensitive to the values of  the non-Abelian fluxes. In the Abelian regime $\Phi_\alpha=\Phi_\beta=0$, we observe that the Hall conductivity  follows the usual integer QHE $\sigma_{xy}=\frac{2 \nu}{h}$, where the factor 2 is due to color-degeneracy. Conversely, in the $\pi$-flux regime ($\Phi_\alpha=\Phi_\beta=\pi/2$) illustrated in Fig.\ref{thefig2}.b, we obtain a completely different sequence of Hall plateaus where $\sigma_{xy}=\frac{4}{h}(\nu+\frac{1}{2})$ around $E_{\text{F}}=0$, as recently observed in graphene \cite{nature2}. This sequence is characterized by sudden changes of sign across the VHS situated at $E= \pm 2$, and by unusual double steps which can be traced back to the underlying low-energy relativistic  excitations.  As the gauge fluxes vary in the vicinity of the $\pi$-flux point ($\Phi_\alpha=\pi/2+\epsilon$ and $\Phi_\beta=\pi/2-\epsilon$), the system enters the non-Abelian regime and the Hall plateaus are modified (see Fig.\ref{thefig2}.d). Indeed most of the degeneracies induced by the Dirac points are  lifted and the anomalous double steps around $E_{\text{F}}=0$ are progressively destroyed. However, a striking behavior occurs: as the non-Abelian fluxes are varied, the two VHS originally situated at $E = \pm 2$ in the $\pi$-flux point are split into four 
\begin{equation}
E^{\txt{VHS}}_{\txt{red}}=  \pm 2 (1 + \cos \Phi_\beta), \hspace{1ex} E^{\txt{VHD}}_{\txt{green}}=  \pm 2 (1 + \cos \Phi_\alpha),
\label{vhs}
\end{equation}
as  illustrated in Fig.\ref{thefig2}.c for $\epsilon=0.1$. Surprisingly enough, anomalous double steps in the plateau sequence reappear at higher energies outside the two \emph{red} VHS, while the \emph{green} VHS induce a sudden change of sign (see Figs.\ref{thefig2}.c-d). It is interesting to note  that the anomalous behavior persists in the high-energy regime and that this effect can be probed by varying the parameter $\Phi_{\beta}$. The temperature required to observe these plateaus should be smaller than the spectral gaps, namely $T\sim 10$  nK.

To identify the  non-Abelian features in this QHE, we introduce the Abelian flux $\Phi$ in the  Dirac Hamiltonian~\eqref{dirac_equation}  by minimal coupling $\textbf{p}\to\textbf{p}+\textstyle{\frac{B_0}{2}}(-y,x)$, and obtain
\begin{equation}
\label{JC_AJC_hamiltonian}
H_{\text{D}}=\left(g_-\sigma^+a+g_-\sigma^-a^{\dagger})+(g_+\sigma^+a^{\dagger}+g_+\sigma^{-}a\right),
\end{equation}
where    $\sigma^+=\ket{\chi_{1}}\bra{\chi_{2}}$, $\sigma^-=\ket{\chi_{2}}\bra{\chi_{1}}$ are  color-flip operators,   $g_{\pm}=(c_y\pm c_x) ( B_0/2)^{1/2}$, and $a^{\dagger},a$ are bosonic chiral operators listed in~\cite{comment}. In the isotropic limit $g_-=0$, the Hamiltonian consists of an anti-Jaynes-Cummings  term, a well-known interaction in quantum optics~\cite{JC} that leads to the usual relativistic Landau levels (LL) recently observed in graphene \cite{graphene_review}. Conversely, in the non-Abelian regime $g_-\neq0$, the Hamiltonian becomes a simultaneous combination of Jaynes-Cummings  and anti-Jaynes-Cummings  terms, producing  a new type of  Landau levels. These novel LL are obtained by means of a Bogoliubov squeezing transformation $S(\zeta)=\ee^{\frac{\zeta}{2}(a^2-(a^{\dagger})^2)}$ with  $\zeta= - \text{tanh}^{-1}(g_-/g_+)$, leading to the energy spectrum
\begin{equation}
\label{rll}
E_{\text{LLL}}=0,\hspace{2ex}E_{n}^{\pm}=\pm\sqrt{(2  B_0c_xc_y)n}, \hspace{2 ex} n=1,2...
\end{equation} 
and corresponding eigenstates 
\begin{equation}
\begin{split}
\ket{\text{LLL}}&=\ket{\chi_2}S^{\dagger}(\zeta)\ket{\text{vac}},\\
\ket{E_{n}^{\pm}}\hspace{1ex}&=\textstyle{\frac{1}{\sqrt{2}}}\ket{\chi_1}S^{\dagger}(\zeta)\ket{n-1}\pm \textstyle{\frac{1}{\sqrt{2}}}\ket{\chi_2}S^{\dagger}(\zeta)\ket{n},
\end{split}
\end{equation}
with  $\ket{n}=(n!)^{-1/2}(a^{\dagger})^{n}\ket{\text{vac}}$ being the usual Fock states. 
Accordingly, the effect of non-Abelian fields is to squeeze the usual LL. In particular, the lowest Landau level (LLL) is a zero-energy mode characterized by a  colored squeezed vacuum, which is in clear contrast with its Abelian counterpart,   the latter being simply the vacuum. Besides, this LLL presents half the degeneracy of the remaining excited states $n\geq1$~\cite{sharapov}, and leads to the so-called anomalous half-integer QHE 
\begin{equation}
\label{Hall}
\sigma_{xy}=\textstyle{\pm\frac{g}{h}\left(\nu+\half\right)},
\end{equation}
where the filling factor $\nu$ is defined  as the integer part  of $[E_{\text{F}}^2/2  B_0c_xc_y]$, and $g$ is the Dirac points degeneracy. Let us stress that the non-Abelian fluxes modify the Hall plateaus in a non-trivial manner as already emphasized through the numerical results. In particular, the Hall conductivity in Eq.~\eqref{Hall} predicts the anomalous half-integer plateaus represented in Fig.~\ref{thefig2}(b), where the conical singularities are four-fold degenerate $g=4$. Conversely, in the non-Abelian case shown in Fig.~\ref{thefig2}(d), the degeneracy is lifted to $g=1$, and thus the size of the steps is modified in accordance. 

\begin{center} 
\begin{figure}
\begin{center}
\hspace{-0.cm}{\scalebox{0.16}{\includegraphics{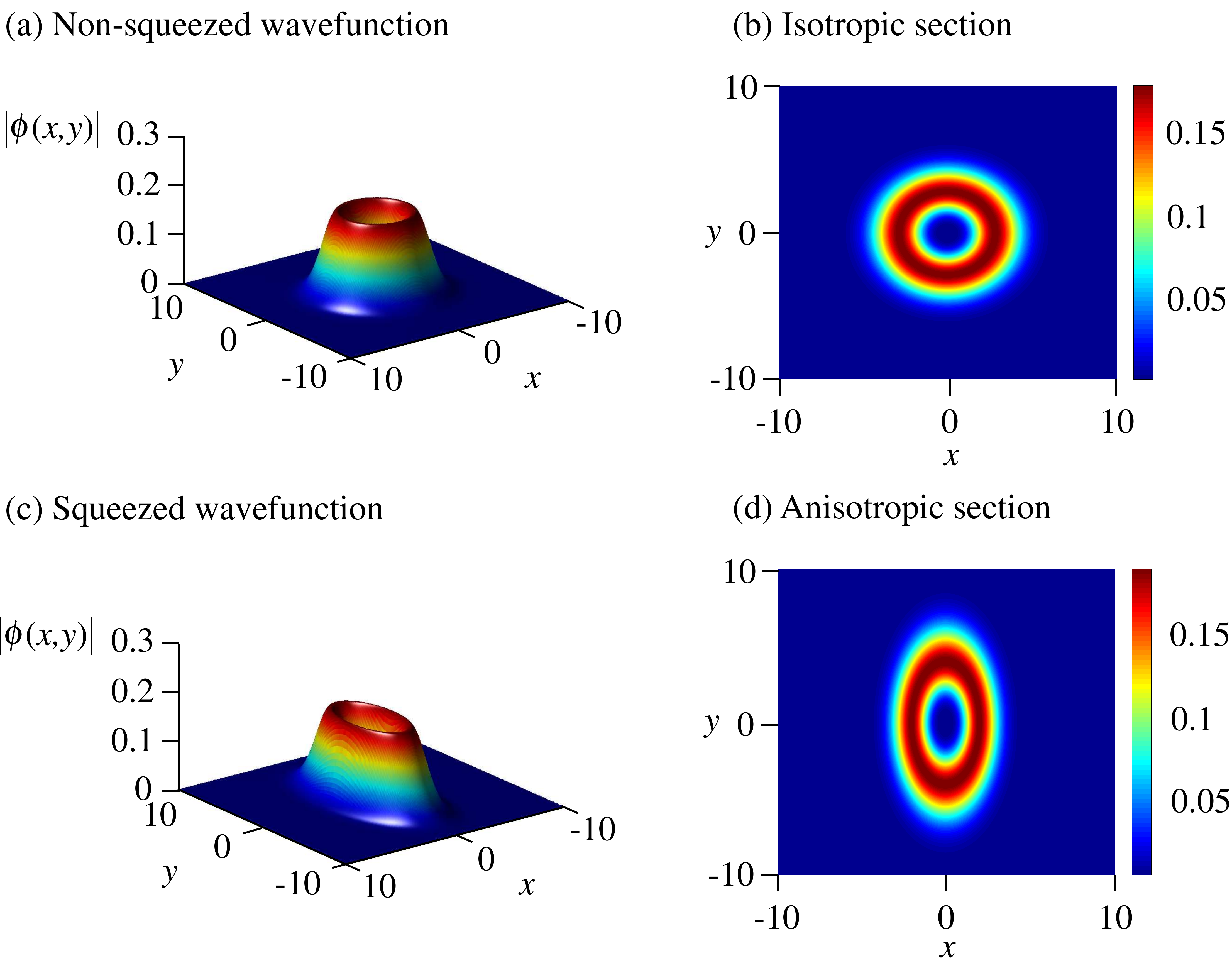}}}
\caption{\label{alejandro} Vortex-like single-particle wavefunctions of the LLL $\phi^{m}_{\text{LLL}}(x,y)$ for $m=4$. (a),(b) Isotropic limit $c_x=c_y$. (c),(d) Anisotropic regime $c_y= 2c_x$. Note that distances are measured in units of the magnetic length $l_{\text{B}}$.} 
\end{center}
\end{figure} 
\end{center}

As discussed above, the anomalous QHE is essentially a single-particle phenomenon that relies on the peculiar properties of the LLL. Additionally, further non-Abelian anomalies can also be found at the many-particle level, where an exotic Laughlin wavefunction~\cite{laughlin} can be obtained  by  filling the  single-particle vortex  wavefunctions
\begin{equation}
\label{LLL_wf}
\phi^{m}_{\text{LLL}}(x,y)=\left(\textstyle{\sqrt{\frac{c_y}{c_x}}x-\ii\sqrt{\frac{c_x}{c_y}}y}\right)^{m}\ee^{-\left(\frac{x^2}{2\tau_x^2}+\frac{y^2}{2\tau_y^2}\right)}.
\end{equation}
Here $\tau_x=l_{\text{B}}\sqrt{2c_x/c_y}$, $\tau_y=l_{\text{B}}\sqrt{2c_y/c_x}$, describe the anisotropic extent of the wavefunction in units of the magnetic length $l_\text{B}=\sqrt{1/B_0}$, and $m=0,1...$ represents the number of left-handed quanta~\cite{comment}. Note how the loss of rotational invariance caused by the non-Abelian induced anisotropy $c_x\neq c_y$,  leads to the squeezing of the vortex levels (Figs.~\ref{alejandro}(a)-(d)).  Filling these squeezed degenerate  states~\eqref{LLL_wf} according to  Fermi statistics, we obtain the Laughlin wavefunction 
\begin{equation}
\label{squeezed_laughlin}
\Psi[z]=\prod_{j<k}(uz_{jk}-v\bar{z}_{jk})\ee^{-\sum_{j}f(u,v)|z_j|^2-g(u,v)(z_j^2+\bar{z}_j^2)},
\end{equation}
where $u=\text{cosh} \zeta$,  $v=\text{sinh} \zeta$, $f(u,v)=\textstyle{\frac{1}{4}}(u^2+v^2)$ and $g(u,v)=\textstyle{\frac{1}{4}}uv$ depend on the anisotropy through the squeezing parameter $\zeta$, and  $z_{jk}=z_j-z_k$ represents the complex two-fermion distance. In the Abelian limit $\zeta=0$, one recovers the standard integer Laughlin wavefunction $\Psi[z]=\prod_{j<k}f(z_j,z_k)\ee^{-\sum_{j}|z_j|^2/4l_{\text{B}}^2}$, where $f(z_j,z_k)=z_j-z_k$ belongs to the space of holomorphic functions (Bargman-Fock space~\cite{girvin}). Strikingly, in the non-Abelian scenario $\zeta\neq 0$, the  wavefunction \eqref{squeezed_laughlin} does not belong to such space due to the interference between holomorphic $f(z)$ and antiholomorphic $f(\bar{z})$ components, and thus represents an instance of a non-chiral QHE. As shown below, this new anomaly modifies the classical analogy with the  one-component plasma (OCP), the
building block that characterizes the peculiar properties of quasiparticles in the fractional QHE~\cite{hall_book}. The Laughlin state can be interpreted as the partition function of a  OCP
$|\Psi[z]|^2\propto Z_{\text{c}}=\int\prod_j dz_jd\bar{z}_j \ee^{-U_{\text{c}}/kT}$ with $kT=\textstyle{1/2}$, a classical gas of particles interacting with a charged background through the potential
\begin{equation}
\label{plasma}
U_{\text{c}}=-\sum_{jk}\text{log}\big| uz_{jk}-v\bar{z}_{jk}\big|+\textstyle{\frac{1}{4}}\sum_{j}\left(f|z_j|^2-g(z_j^2+\bar{z}_j^2)\right).
\end{equation} 
The last term corresponds to the charged background jellium $\rho_{\text{j}}=-\frac{1}{4\pi l_{\text{B}}^2}(\frac{c_x}{c_y}+\frac{c_y}{c_x})$, whereas the first  describes a collection of positively charged particles $q=1$ surrounded by a charge cloud $\delta\rho(z)$,  with $z=|z|\ee^{-\ii\theta}$, and
\begin{equation}
\delta\rho(|z|,\theta)=\frac{\text{tanh}\zeta}{|z|^2}\frac{(1+\text{tanh}^2\zeta)\cos2\theta-4\text{tanh}\zeta}{(1+\text{tanh}^2\zeta)-4\text{tanh}\zeta \cos2\theta}.
\end{equation}
Notice how the surrounding charge cloud is absent $\delta\rho(z)=0$ in the Abelian limit $\zeta=0$, and we recover the usual OCP analogy. Conversely, for non-Abelian regimes, the  collection of  interacting positively charged particles becomes locally surrounded by an anisotropic charge cloud $\rho=\sum_jq\delta(z-z_j)+\delta\rho(z_j)$ with $\int d^2z\delta\rho(z)=0$. In accordance, the paradigmatic plasma analogy is altered due to the squeezed nature of the LLL, a fact that may find profound consequences in the fractional QHE. \\

We have shown that non-Abelian optical lattices offer an intriguing route to probe the striking properties of emerging Dirac fermions in anisotropic Minkowski space-times. In particular, the versatility offered by such experimental setups leads to the unique possibility of tuning the anisotropy of the underlying space-time, leading to remarkable effects such as non-chiral  quantum Hall effects with several types of anomalies.\\

We acknowledge the support of ERC AdG QUAGATUA, EU IP SCALA, EU STREP NAMEQUAM m ESF/Spanish MEC Euroquam Programm  FERMIX, MEC Grant TOQATA, the Belgian Federal Government, the ``Communaut\'e fran\c caise de Belgique", F.R.S.-FNRS, FIS2006-04885, the Polilsh Government Scientific Funds 2009-10, CAM-UCM/910758, INSTANS 2005-2010, FPU MEC grant.
We thank J. Schliemann for discussion. 
%%%%%%%%%%%%%%%%%%%%%%%%%%%%%%%%%%%%%%%%%%%%%%%%%%%%%%%%%%%%%%%%%%%

%%%%%%%%%%%%%%%%%%%%%%%%%%%%%%%%%%%%%%%%%%%%%%%%%%%%%%%%%%%%


\begin{references}
\bibitem{comment}
Right-handed  $a=(a_x-\ii a_y)/\sqrt{2}$ and left-handed  $b=(a_x+\ii a_y)/\sqrt{2}$ operators are defined in terms of the usual cartesian modes $a_{j}=\sqrt{\frac{\omega}{2}}(x_j+ \frac{\ii}{\omega}p_j)$, where $\omega=B_0/2$.
\end{references}
\end{document}